\documentclass{pasj00}
\usepackage{color}

\begin{document}
\SetRunningHead{Y. Takeda et al.}{CNO abundances of A-type stars }
\Received{2018/05/02}
\Accepted{2018/07/16}

\title{Photospheric carbon, nitrogen, and oxygen abundances \\
of A-type main-sequence stars 
\thanks{Based on observations carried out at Okayama Astrophysical Observatory
(National Astronomical Observatory of Japan) and  
Bohyunsan Astronomical Observatory (Korean Astronomy and Space Science Institute).}
}

\author{
Yoichi \textsc{Takeda},\altaffilmark{1,2}
Satoshi \textsc{Kawanomoto},\altaffilmark{1}
Naoko \textsc{Ohishi},\altaffilmark{1}\\
Dong-Il \textsc{Kang},\altaffilmark{3}
Byeong-Cheol \textsc{Lee},\altaffilmark{4,5}
Kang-Min \textsc{Kim},\altaffilmark{4,5}
and
Inwoo \textsc{Han}\altaffilmark{4,5}
}
\altaffiltext{1}{National Astronomical Observatory, 2-21-1 Osawa, Mitaka, Tokyo 181-8588, Japan}
\email{takeda.yoichi@nao.ac.jp, kawanomoto.satoshi@nao.ac.jp, naoko.ohishi@nao.ac.jp}
\altaffiltext{2}{SOKENDAI, The Graduate University for Advanced Studies, 
2-21-1 Osawa, Mitaka, Tokyo 181-8588}
\altaffiltext{3}{Changwon Science high school,
30, Pyungsanro 159-th, Uichang, Changwon, 641-500, Korea}
\email{kangdongil@gmail.com}
\altaffiltext{4}{Korea Astronomy and Space Science Institute,
776, Daedeokdae-Ro, Youseong-Gu, Daejeon 34055, Korea}
\email{bclee@kasi.re.kr, kmkim@kasi.re.kr, iwhan@kasi.re.kr}
\altaffiltext{5}{Korea University of Science and Technology, 217, 
Gajeong-ro Yuseong-gu, Daejeon 34113, Korea}



\maketitle

\begin{abstract}
Based on the spectrum fitting method applied to C~{\sc i}~5380, N~{\sc i}~7486, 
and O~{\sc i}~6156--8 lines, we determined the abundances of C, N, and O  
for 100 mostly A-type main-sequence stars (late B through early F at 
11000~K~$\gtsim T_{\rm eff} \gtsim$~7000~K) comprising normal stars as well as 
non-magnetic chemically peculiar (CP) stars in the projected rotational velocity
range of 0~km~s$^{-1}$~$\ltsim v_{\rm e}\sin i \ltsim$~100~km~s$^{-1}$, 
where our aim was to investigate the abundance anomalies of these elements
in terms of mutual correlation, dependence upon stellar parameters, and
difference between normal and CP stars.
We found that CNO are generally underabundant (relative to the standard star 
Procyon) typically by several tenths dex to $\sim 1$~dex for almost all stars 
(regardless of CP or normal), though those classified as peculiar (Am or HgMn) 
tend to show larger underabundance, especially for C in late Am stars and for 
N in HgMn stars of late B-type, for which deficiency amounts even up to $\sim 2$~dex.
While the behaviors of these three elements are qualitatively similar to each other,
the quantitative extent of peculiarity (or the vulnerability to the physical 
process causing anomaly) tends to follow the inequality relation of C~$>$~N~$>$~O.
Regarding the considerable star-to-star dispersion observed at any $T_{\rm eff}$,
the most important cause is presumably the difference in rotational velocity. 
These observational facts appear to be more or less favorably compared 
with the recent theoretical calculations based on the model of atomic 
diffusion and envelope mixing. 
\end{abstract}

%


\section{Introduction}

It has long been known that a fraction of stars in the upper main sequence  
around A-type show unusual spectra characterized by conspicuous strengthening 
of specific metallic lines (or weakening of some lines such as Ca~{\sc ii} K),
indicating that the abundances of relevant elements are anomalous in their atmosphere.
These stars are thus called ``chemically peculiar stars'' (CP stars), which are 
further divided into several subclasses (e.g., Preston 1974).
Since the notable property of these CP stars that they are generally sharp-lined 
and thus rotate more slowly in comparison with normal stars, slow rotation is 
considered to be the important factor for producing abundance anomalies. 
Theoretically, the most promising mechanism for the origin of such chemical 
peculiarity is ``atomic diffusion''; i.e., a kind of element segregation process 
caused by imbalance between the downward gravitational force and the upward radiative 
force, which effectively acts when the atmosphere/envelope is sufficiently stable 
to prevent from substantial mixing (see, e.g., Michaud, Alecian, \& Richer 2015;
and the references therein). In this scenario, slow rotation is regarded as 
a necessary condition to guarantee the stability required for the build-up of anomaly.
 
Among the elements involved in CP phenomena, the light elements C, N, and O
(which are most abundant next to H and He and thus of profound astrophysical 
importance) exhibit rather unique characteristics, in the sense that they  
show deficiencies unlike many other heavier species tending to suffer 
overabundances, because upward radiative acceleration is not so large as to 
counter the downward gravity acceleration (e.g., Gonzalez, Artru, \& Michaud 1995).
However, since the equilibrium surface abundances are eventually determined 
by how the hydrodynamical mixing process (e.g., meridional circulation, 
turbulence, mass loss, etc.) actually operates in the stellar envelope
or atmosphere, the theoretical prediction heavily depends on the modeling
details (e.g., how the mixing-related parameters are chosen), the validity 
of which has to be empirically checked in comparison with observations.

Unfortunately, previous observational studies of CNO abundances in A-type 
stars are not necessarily satisfactory in this respect, despite that 
a large number of spectroscopic investigations on the chemical abundances 
of CP and normal A stars have been published so far. \\
--- First, the size and diversity of the sample is generally not sufficient, 
because such spectroscopic study published in a paper usually targets only 
a comparatively small number of stars. Moreover, they tend to be biased toward 
sharp-line stars of small $v_{\rm e}\sin i$ (projected rotational velocity) 
because of the increasing difficulty in analyzing the spectra of broad-line stars. 
In order to provide useful observational constraints for theoretical modeling of 
atomic diffusion, sample stars are desired to cover a wide range of 
relevant stellar parameters, which are considered to affect the prediction
of surface abundance peculiarity (e.g., $v_{\rm e}\sin i$, $T_{\rm eff}$). \\
--- Second, although a number of previous investigations (e.g., Fossati et al. 2007, 2009;
Royer et al. 2014; and the references therein) included any of CNO
(especially C or O) along with various other elements, only a few have focused 
on consistently studying the abundances for all the three in A-type stars  
and comparing them with each other (e.g., Roby \& Lambert 1990; Leushin et al. 1992;
Savanov 1995). Actually, the pioneering work by Roby and Lambert (1990),
who studied the CNO abundances of 37 CP stars and 5 standard A and late-B stars,
still remains as the most widely quoted study in this field.
Especially, information of N abundances for such A-type and related stars 
is evidently insufficient, presumably because N lines generally suffer 
considerable non-LTE effect (Takeda 1992b; Rentzsch-Holm 1996;  Lemke \& Venn 1996;
Przybilla \& Butler 2001). This situation should be redressed by all means. 

Motivated by this circumstance, we decided to conduct a comprehensive study 
on the abundances of C, N, and O for a large sample of 100 late-B through early-F
stars (including not only CP stars but also normal stars) covering sufficiently
large ranges of stellar parameters (11000~K~$\gtsim T_{\rm eff} \gtsim$~7000~K
and 0~km~s$^{-1}$~$\ltsim v_{\rm e}\sin i \ltsim$~100~km~s$^{-1}$),
where the synthetic spectrum-fitting method was applied (inevitable for 
analyzing the spectra of broad-line stars) and the non-LTE effect was properly 
taken into consideration. This is kind of an extension of our previous work 
(Takeda et al. 2008, 2009; hereinafter referred to as T08 and T09, respectively), 
in which C or O were included as target elements), but the sample has been 
considerably refined by adding new observational data and attention has been 
paid also to N. The points of interest which we want to clarify are as follows: 
\begin{itemize}
\item
Do the abundance trends of these three elements correlate well with each other? 
Or do they show any notable difference? What about their relation to the metallicity (Fe)? 
\item
How do the extents of peculiarity in C, N, and O depend upon the stellar parameters?
Can we observe any dependence upon $v_{\rm e}\sin i$ or $T_{\rm eff}$?
\item
Are there any distinct difference between CP stars and normal stars 
in terms of CNO abundances? Do stars classified as normal exhibit 
any sign of chemical peculiarity?   
\end{itemize}

\section{Observational data}

Given the spectral data of a large number of stars of B--A--F types
currently available to us (either the data already used in our previous 
studies or newly observed data), we selected our target sample 
by considering the following requirements:\\
--- As one of our main aims was to examine the effect of 
stellar rotation on CNO abundance peculiarities, we would like to 
include stars in as wide $v_{\rm e}\sin i$ range as possible.
However, according to our experience (T08, T09), abundance determination
for very rapid rotators (e.g., $v_{\rm e}\sin i \sim $~200--300~km~s$^{-1}$)
is not easy and often fails, while our spectrum-fitting approach turned out 
successful for most stars of $v_{\rm e}\sin i \ltsim $~100~km~s$^{-1}$. 
Since Abt and Morrell (1995) reported that abundance anomalies are 
seen mostly for stars with $v_{\rm e}\sin i$ lower than $\sim 120$~km~s$^{-1}$, 
we decided to confine our targets only to those 
of $v_{\rm e}\sin i \le $~100~km~s$^{-1}$.\\
--- Although our main emphasis is placed on A-type stars, 
they had better be discussed in company with early-F stars (where 
Am-like anomaly is also observed) and late-B stars (where HgMn peculiarity 
is seen), which are considered to be closely related.\\
--- Among the various types of CP stars on the upper main sequence, we 
already learned that spectra of magnetic CP stars (in which variability 
is often involved) tend to be so complex that they can not be well fitted 
by the conventional spectrum modeling. Therefore, we concentrated only 
on non-magnetic CP stars (Am stars or HgMn stars).\\
--- In order to investigate the cause of abundance peculiarity,  
studying a sufficient number of cluster stars altogether is useful,
because they should have been born with the same initial composition.
For this purpose, we included A--F stars belonging to the Hyades cluster.
 
Consequently, our program stars consist of 100 mostly A-type stars 
(late B through early F) on or near to the main sequence (luminosity classes 
of III--V) which have slow to moderately-high rotational velocities
(0~km~s$^{-1}$~$\ltsim v_{\rm e}\sin i \ltsim 100$~km~s$^{-1}$). 
Among these, about $\sim 30$\% are non-magnetic CP stars: 25 Am stars 
(from the spectral type given in the Hipparcos catalogue; cf. ESA 1997) 
and 5 HgMn (or Mn) stars (cf. table~1 of Takeda et al. 1999). Besides, 
16 stars (about $\sim 1/6$) are Hyades stars.
The list of these program stars is given in table~1.
It may be worth noting that a significant fraction of our targets
are spectroscopic binaries (data taken from Hoffleit \& Jaschek 1991), 
for which the high binary frequency in Am stars (cf. Preston 1974) may be 
at least partially responsible. However, apparently doubled-lined 
binaries (such as those difficult to be modeled by the theoretical
spectrum simulated for a single star) are not included in our sample stars. 

Regarding the observational data, we could avail ourselves of the spectra 
already at our hand for 71 stars and the standard star Procyon, which were 
used in our previous studies, as summarized in table~2.
Meanwhile, the data of 29 stars were secured by our new observations 
carried out on 2017 August 22--23 and November 6 by using the 188~cm reflector 
along with HIDES (HIgh Dispersion Echelle Spectrograph)  
at Okayama Astrophysical Observatory.
The data reduction was done in the standard manner by using IRAF,\footnote{
  IRAF is distributed by the National Optical Astronomy Observatories,
  which is operated by the Association of Universities for Research
  in Astronomy, Inc. under cooperative agreement with
  the National Science Foundation.} which resulted in spectra with the resolving 
power of $R\sim 100000$ covering the wavelength range of 5100--8800~$\rm\AA$.
For most of the spectra of our 100 targets, sufficiently high S/N ratios 
(typically on the order of $\sim 200$) are attained.

\section{Stellar parameters and model atmospheres}  

As in our previous studies (see the references given in table~2),
the effective temperature ($T_{\rm eff}$) and the surface gravity ($\log g$) 
for each of the 100 program stars were determined from the colors of 
Str\"{o}mgren's $uvby\beta$ photometric system with the help of 
Napiwotzki, Sch\"{o}nberner, and Wenske's (1993) 
{\tt uvbybetanew} program\footnote{
$\langle$http://www.astro.le.ac.uk/\~{}rn38/uvbybeta.html$\rangle$.}, 
where the observational data of $b-y$, $c_{1}$, $m_{1}$, and $\beta$ 
were taken from Hauck and Mermilliod (1998) via the SIMBAD database.
Their typical errors may be on the order of $\sim 3\%$ in 
$T_{\rm eff}$ and $\sim 0.1$~dex in $\log g$ for the present
case of stars around A-type (cf. Sect.~5 of Napiwotzki et al. 1993).
Regarding the microturbulence ($v_{\rm t}$), we adopted the analytical 
$T_{\rm eff}$-dependent relation derived in T08
\begin{equation}
v_{\rm t} = 4.0 \exp\{- [\log (T_{\rm eff}/8000)/A]^{2}\} \\
\end{equation}
(where $A \equiv [\log (10000/8000)]/\sqrt{\ln 2}$),
which roughly represents the observed distribution of $v_{\rm t}$
with probable uncertainties of $\pm 30\%$ (cf. Fig.~2b in T08).
The only exception is the standard star Procyon, for which 
we used Takeda et al.'s (2005b) spectroscopically determined values
($T_{\rm eff}$ = 6612~K, $\log g = 4.00$, and $v_{\rm t} = 2.0$~km~s$^{-1}$)
in order to maintain consistency with T08.
The adopted values of $T_{\rm eff}$, $\log g$, $v_{\rm t}$ are summarized in table~1.
All the program stars are plotted on the $\log L$ vs. $\log T_{\rm eff}$ 
diagram (theoretical HR diagram) in figure~1, where seven representative
theoretical evolutionary tracks corresponding to different stellar 
masses are also depicted. We can see from this figure that the masses 
of our sample stars are in the range between $\sim 1.5 M_{\odot}$ and 
$\sim 5 M_{\odot}$. More detailed  data regarding the targets and 
their stellar parameters are given in the electronic table (tableE.dat) 
presented as the online material.

The model atmosphere for each star was then constructed by two-dimensionally 
interpolating Kurucz's (1993) ATLAS9 model grid (for $v_{\rm t} = 2$~km~s$^{-1}$) 
in terms of $T_{\rm eff}$ and $\log g$, where the solar-metallicity models 
were exclusively used as in our previous studies. We also computed the non-LTE
departure coefficients for C, N, and O corresponding to each atmospheric model,
which are to be used for non-LTE abundance analysis, by following the procedure 
described in Takeda (1992b) (for C and N) and Takeda (1992a, 2003) for O. 

\section{Determination of CNO abundances} 

Following our previous studies, we invoke the C~{\sc i} 5380.337 line 
(as in T08) and the O~{\sc i} 6156--8 feature comprising 9 components 
(as in T08, T09, Takeda et al. 2012, 2014) for deriving the C and O abundances.
As to N, we decided to adopt the N~{\sc i} line at 7468.312~\AA, which 
has a suitable strength as an abundance indicator. 
The determination procedures of abundances and related quantities (e.g., 
non-LTE correction, uncertainties due to ambiguities of atmospheric parameters) 
are essentially the same as in our previous papers quoted above, which consist 
of two consecutive steps.

\subsection{Synthetic spectrum fitting} 

The first step is to find the solutions for the abundances of relevant elements 
($A_{1}, A_{2}, \ldots$), projected rotational velocity ($v_{\rm e}\sin i$), 
and radial velocity ($V_{\rm rad}$) such as those accomplishing the
best fit (minimizing $O-C$ residuals) between theoretical and 
observed spectra, while applying the automatic fitting algorithm 
(Takeda 1995). Three wavelength regions were selected for this purpose:
(1) 5375--5390~\AA\ region (for C), (2) 7457--7472~\AA\ region (for N), and 
(3) 6146--6163~\AA\ region (for O). More information about this fitting analysis 
(varied elemental abundances, used data of atomic lines) is summarized in table~3.
How the theoretical spectrum for the converged solutions fits well 
with the observed spectrum is displayed in figures~2--4 for each region. 
The $v_{\rm e}\sin i$ values\footnote{It should be 
kept in mind that we assumed only the rotational broadening (with the 
limb-darkening coefficient of $\epsilon = 0.5$) as the macrobroadening 
function to be convolved with the intrinsic theoretical line profiles.
Accordingly, $v_{\rm e}\sin i$ values for very sharp-line cases 
(e.g., $v_{\rm e} \sin i \ltsim$~5--6~km~s$^{-1}$) 
should be regarded rather as upper limits because the effects of 
instrumental broadening and macroturbulence are neglected.
} resulting from the fitting of 
6146--6163~\AA\ region are presented in table~1. We also adopted the
solution of Fe abundance derived from the fitting of 6146--6163~\AA\ region 
as the metallicity of each star (given as [Fe/H] in table~1).  

\subsection{Abundances from equivalent widths} 

As the second step, with the help of Kurucz's (1993) WIDTH9 program 
(which had been considerably modified in various respects; e.g., 
inclusion of non-LTE effects, treatment of total equivalent width for 
multi-component lines; etc.), we computed the equivalent widths ($W$) 
of the representative lines ``inversely'' from the abundance solutions
(resulting from spectrum synthesis) along with the adopted atmospheric 
model/parameters; i.e., $W_{5380}$ (for C~{\sc i} 5380), 
$W_{6156-8}$ (for O~{\sc i} 6156--8), and $W_{7468}$ (for N~{\sc i} 7468),
because they are easier to handle in practice (e.g., for estimating the
uncertainty due to errors in atmospheric parameters).
The adopted atomic data for these lines are summarized in table~4.  
We then analyzed such derived $W$ values by using WIDTH9 to determine 
$A^{\rm N}$ (NLTE abundance) and $A^{\rm L}$ (LTE abundance), from 
which the NLTE correction $\Delta (\equiv A^{\rm N} - A^{\rm L})$
was further derived. Since we adopted Procyon as the standard star of
abundance reference, which is known to have essentially the same abundance
as the Sun (cf. Sect.~IV(c) in T08; see also Takeda 1994), we define 
the relative abundance as [X/H] $\equiv$ $A_{\rm X}^{\rm N}$(star) $-$ 
$A_{\rm X}^{\rm N}$(Procyon) (X = C, N, O). The resulting values of 
[C/H], [N/H], and [O/H] are given in table~1 (more complete results
including $W$ and $\Delta$ are separately presented in ``tableE.dat''). 
Figures 5(C), 6(N), and 7(O) graphically show the equivalent width ($W$),
non-LTE correction ($\Delta$), non-LTE abundance ($A^{\rm N}$), 
and abundance variations in response to parameter changes (see the 
following subsection~4.3), as functions of $T_{\rm eff}$. 
As we can recognize in panel (b) of these figures, while the non-LTE corrections 
for C~{\sc i} 5380 and O~{\sc i} 6156--8 are not very significant 
($|\Delta| \ltsim 0.1$~dex), those for N~{\sc i} 7486 are rather important 
(up to $|\Delta| \ltsim$ 0.4--0.5~dex) and thus have to be taken into 
consideration especially at higher $T_{\rm eff}$.\\

\subsection{Error estimation}

In order to evaluate the abundance errors caused by uncertainties
in atmospheric parameters, we estimated six kinds of abundance variations
($\delta_{T+}$, $\delta_{T-}$, $\delta_{g+}$, $\delta_{g-}$, 
$\delta_{v+}$, and $\delta_{v-}$) for $A^{\rm N}$ by repeating the 
analysis on the $W$ values while 
perturbing the standard atmospheric parameters interchangeably by 
$\pm 3\%$ in $T_{\rm eff}$, $\pm 0.1$~dex in $\log g$, 
and $\pm 30\%$ in $v_{\rm t}$ (which are the typical 
ambiguities of the parameters we adopted; cf. section~3). 
Finally, the root-sum-square of these perturbations,
$\delta_{Tgv} \equiv (\delta_{T}^{2} + \delta_{g}^{2} + \delta_{v}^{2})^{1/2}$, 
were regarded as abundance uncertainties (due to combined errors in 
$T_{\rm eff}$, $\log g$, and $v_{\rm t}$), 
where $\delta_{T}$, $\delta_{g}$, and $\delta_{\xi}$ are defined as
$\delta_{T} \equiv (|\delta_{T+}| + |\delta_{T-}|)/2$, 
$\delta_{g} \equiv (|\delta_{g+}| + |\delta_{g-}|)/2$, 
and $\delta_{v} \equiv (|\delta_{v+}| + |\delta_{v-}|)/2$,
respectively. 
These $\delta_{T\pm}$, $\delta_{g\pm}$, and $\delta_{v\pm}$ are plotted
against $T_{\rm eff}$ in panels (d), (e), and (f) of figures~5--7.
We can see that only $\delta_{T\pm}$ can be appreciably significant ($\ltsim$~0.1--0.2~dex)
reflecting the high-excitation nature of the adopted lines, while $\delta_{g\pm}$ 
and $\delta_{v\pm}$ are of negligible importance (insensitivity to changing 
$v_{\rm t}$ is interpreted as due to the large thermal velocity for
these light elements).

We also evaluated errors due to random noises of the observed spectra by 
estimating S/N-related uncertainties in the equivalent width ($W$) 
by invoking the relation derived by Cayrel (1988),
$\delta W \simeq 1.6 (w \delta x)^{1/2} \epsilon$,
where $\delta x$ is the pixel size (0.03~$\rm\AA$), $w$ is the full-width at half 
maximum (which may be roughly regarded as  $\sim \lambda v_{\rm e}\sin i/c$; 
where $\lambda$ is line wavelength and $c$ is the velocity of light), 
and $\epsilon \equiv ({\rm S/N})^{-1}$; typically $\sim 1/200$).
Then, we determined the abundances for each of the perturbed $W_{+} (\equiv W + \delta W)$ 
and $W_{-} (\equiv W - \delta W)$, from which the differences from the standard 
abundance ($A$) were derived as $\delta_{W+} (>0)$ and $\delta_{W-} (<0)$.
We thus regard $\delta_{W} \equiv 0.5 (\delta_{W+} + |\delta_{W-}|)$ as the
abundance uncertainties due to photometric noises. 
Since the equivalent width error ($\delta W$) is in the range of only $\sim$~0.3--2~m\AA,
the corresponding abundance ambiguity ($\delta_{W}$) is generally insignificant 
(on the order of a few hundredths dex in most cases).

Exceptionally, however, in case of very weak lines, $\delta_{W}$ can be appreciably 
large as much as several tenths dex. 
As a tentative criterion, we regard that the resulting abundance is unreliable 
if $W$ is smaller than 3$\delta_{W}$.\footnote{We can use $W_{5380}$ and $W_{7648}$
for $W$ as it is. However, since $W_{6156-8}$ is the combined equivalent width of 
the feature apparently seen as a triplet (cf. table~4), we substituted $W_{6156-8}$/3 
for $W$.} We then found that 11 [C/H] values and 4 [N/H] values (none of the [O/H] values)
satisfy this condition, which are thus unreliable and should be viewed with caution. 
Practically, they had better be regarded rather as upper limits.
These results of large uncertainties are shown with parentheses in table~1, 
and the relevant plots are marked by open circles in figure~5a,c and figure~6a,c.

Finally, combining $\delta_{Tgv}$ and $\delta_{W}$, we obtained
the total error as $\delta_{TgvW} \equiv (\delta_{Tgv}^{2} + \delta_{W}^{2})^{1/2}$,
which are shown as error bars attached to the non-LTE abundances in panel (b)
of figures~5--7.  According the characteristics mentioned above, $\delta_{TgvW}$
is generally dominated by $\delta_{Tgv}$ (i.e., $\delta_{TgvW} \simeq \delta_{Tgv}$), 
excepting the cases of very weak lines where $\delta_{W}$ begins to show 
appreciable/dominant contribution. 
The detailed values of $\delta_{Tgv}$, $\delta_{W}$, and $\delta_{TgvW}$ for each 
star are given in ``tableE.dat'' (available as online material). 

\section{Discussion}

\subsection{Observed behaviors of [C/H], [N/H], and [O/H]} 

The relative abundances ([C/H], [N/H], [O/H], and [Fe/H]) for each star resulting 
from our analysis in section~4 are plotted against $T_{\rm eff}$ and $v_{\rm e}\sin i$
in figure~8 (panels a--h), where their mutual correlations are also shown 
(panels i--n). The following characteristics can be read from these figures.\\
--- C, N, and O are underabundant for almost all program stars (regardless of
whether being classified as peculiar or normal, though larger anomaly tends to 
be seen in the former) typically by several tenths dex to $\sim 1$~dex 
($-1 \ltsim$~[C,N,O/H]~$\ltsim 0$), in contrast to [Fe/H] distributing 
around [Fe/H]~$\sim$~0.\\
--- Moreover, with regard to [C/H] and [N/H], distinctly large deficiencies 
as much as $\sim 2$~dex are shown by a fraction of stars, most of which are CP stars.
That is, especially large depletion is observed in  [C/H] for late Am stars 
($T_{\rm eff} \sim$~7500--8000~K; cf. figure~8e) or in [N/H] for HgMn stars
($T_{\rm eff} \sim$~10000--11000~K; cf. figure~8f). \\
--- When the extents of peculiarity (deficiency) for these three elements
are compared with each other, we may state that (if the conspicuously C- or 
N-depleted stars mentioned above are excluded) the inequality relation
$|[{\rm C}/{\rm H}]| > |[{\rm N}/{\rm H}]| > |[{\rm O}/{\rm H}]|$ 
roughly holds. In other words, carbon is more sensitive than nitrogen and 
nitrogen is more sensitive than oxygen to the mechanism of producing the anomaly. 
It is interesting to note that this is consistent with 
what has been predicted from the diffusion theory (see, e.g., 
Figs.~12--13 of Richer, Michaud, \& Turcotte 2000; 
Figs.~14--16 of Talon, Richard, \& Michaud 2006).\\
--- Regarding the connection between the CNO abundances and the metallicity ([Fe/H]),
we can see a trend of anti-correlation; i.e., [C/H], [N/H], and [O/H] tend to 
decrease with an increase in [Fe/H] (cf. figure~8i, 8j, and 8k). This indicates
that the sense of chemical anomaly acts oppositely for CNO and heavier metals
(cf. section~1).  

\subsection{Dependence upon Stellar Parameters} 

Then, which parameter is most important in determining the extent of peculiarity? 
The $v_{\rm e}\sin i$-dependence of [C/H], [N/H], and [O/H], which is expected
if the mechanism of atomic diffusion (countervailing the rotation-dependent
mixing) is the main cause for the CP phenomenon, is not necessarily clear
in figures~8a, 8b, and 8c. However, this is presumably due to the diversity
of the sample stars (difference of initial composition, etc.). 
In figure~9 are shown the same correlation plots as figure~8 but only for
the selected 16 Hyades stars. We can see from this figure more manifestly 
that [C/H], [N/H], and [O/H] progressively decrease with a decrease
in $v_{\rm e}\sin i$ from $\sim 0$ (at $v_{\rm e}\sin i \sim 100$~km~s$^{-1}$)
to $\sim -1$ (at $v_{\rm e}\sin i \sim 0$~km~s$^{-1}$). 
Similarly to the case of figure~8 mentioned in subsection~5.1, we can recognize 
(much more clearly) the mutual correlation between C, N, and O with the 
inequality relation of $|$[C/H]$| > |$[N/H]$| \gtsim |$[O/H]$|$ 
(cf. panels l, m, and n in figure~9) in this homogeneous sample. 
Actually, Takeda and Sadakane (1997) already  
reported the existence of such trend in Hyades stars for oxygen based on the analysis 
of O~{\sc i} 7771--5 triplet lines. This time, we have confirmed this tendency
not only for O but also for C and N.
It should also be noted that Gebran et al. (2010) reported the tendency of
anticorrelation in [C/H] vs. [Fe/H] as well as [O/H] vs. [Fe/H] plots
(i.e., close correlation between [C/H] and [O/H]) for Hyades A--F dwarfs 
(cf. their Fig.~8) similar to our figures~9i and 9k.

From theoretical point of view, the diffusion model for the explanation of
AmFm peculiarity predicts a $T_{\rm eff}$-dependent tendency; i.e., 
the extent of CNO deficiency progressively increases with decreasing $T_{\rm eff}$
(e.g., figure~14 of Richer et al. 2000). Interestingly, we can recognize
such a trend in figures~8e--8g (and also in figures~9e--9g despite the 
rather narrow $T_{\rm eff}$ range of $\sim$~9000--7000~K) for the combined 
sample\footnote{The [N/H] values in HgMn stars of higher $T_{\rm eff}$ are 
exceptional and should be separately considered, which show conspicuously large 
deficiencies.} of Am stars (red-filled symbols) and normal stars (open symbols). 
That is, the dispersion of [X/H] (X = C, N, O) grows and the lower envelope of
the distribution shifts toward lower values with a decrease in $T_{\rm eff}$.
This may be regarded as being in support of the recent theoretical models 
at least in the qualitative sense.

\subsection{Comparison with previous work}

Finally, we comment on the comparison of our results with several published
studies mentioned in section~1.\\
--- Roby and Lambert (1990) determined the CNO abundances of 13 Am stars, 9 HgMn stars, 
and 5 normal (standard) stars in the $T_{\rm eff}$ range of $\sim$~7000--15000~K,
where most of the sample stars are sharp-lined with $v_{\rm e}\sin i \ltsim 50$~km~s$^{-1}$.
They reported that the mean (C, N, O) abundances of Am stars and HgMn stars
relative to standard stars to be ($-0.2$, $-0.2$, $-0.4$) dex and  
(+0.1, $-0.7$, and $-0.3$) dex. Although these extents of deficiency appear somewhat 
small compared to ours, we should bear in mind that normal stars are not guaranteed to 
have solar CNO composition, as revealed from this study. When we examine the absolute 
abundances (relative to the solar abundances) shown in their Fig.~1 (standard stars), 
Fig~2 (HgMn stars), and Fig.~5 (Am stars), their results are favorably compared
with our figures~8e--8g in the sense that normal stars show marginal underabundances
by $\sim$~0.0--0.3~dex, CNO deficiencies of Am stars have rather large dispersion 
($\sim$~0.0--1.0~dex), and HgMn stars exhibit moderate underabundance (by several 
tenths dex) in C and O along with considerably large deficit ($\gtsim 1$~dex) in N. \\
--- Regarding Leushin et al.'s (1992) analysis on CNO abundances of CP stars,
5 among their 7 targets are of Si or SrCrEu type (magnetic CP stars which are 
not touched in our study) and 2 are HgMn stars.  Only C lines were measured 
for these two HgMn stars to yield [C/H] $\sim -0.4$~dex (reasonably consistent
with our consequence), while N lines were too weak to be measurable (O lines 
were not measured for these stars).\\
--- Savanov (1995) investigated the correlation between the CNO and Fe abundances 
of CP stars and normal stars, based on the already published abundance data
taken from various literature. He reported that [C/H], [N/H], and [O/H] are 
anti-correlated with [Fe/H] (this tendency is still recognized even by excluding 
the weak-line $\lambda$~Boo stars of low [Fe/H], which are out of the scope 
in this study), such as we confirmed in figures~8i--8k and figures~9i--9k.\\
--- Fossati et al. (2007) determined chemical abundances of many elements
for 8 Am stars in the Praesepe cluster (along with 2 normal A-type stars and
1 Blue-Straggler). We see from their Fig.~10 that two normal A-type stars
($T_{\rm eff} \sim$~7400--7800~K) show [C/H]~$\sim 0.0$, [N/H]~$\sim +0.5$,
and [O/H]~$\sim +0.1$ on the average. Regarding Am stars 
($T_{\rm eff} \sim$~7200--8500~K), the abundance ranges are: [C/H] $\sim -0.7$ 
to $0.0$, [N/H] $\sim -0.4$ to $-0.2$ (with an exceptionally 
large value of $\sim +0.4$ for HD~72942), and [O/H] $\sim -0.6$ to $-0.2$. 
Although the general tendency of CNO deficiency for Am stars are surely observed, 
it appears that their abundances are somewhat higher than our results (especially 
for N), which may be due to their neglect of non-LTE corrections.\\  
--- Fossati et al. (2009) carried out elaborate chemical abundance studies 
for 3 sharp-lined normal early-A and late-B stars (HD~145788, 21~Peg, and $\pi$~Cet,
which have $T_{\rm eff}$ of 9750, 10400, and 12800~K, respectively).
Their LTE results show that [C/H], [N/H], and [O/H] for these stars are generally 
supersolar ($>0$) by a few tenths dex (cf. their Fig.~7), which is not consistent 
with our conclusion that even normal stars show some deficits in CNO. Again, this
may indicate that non-LTE effect should be adequately taken into account for deriving 
the abundances of such light elements (as they also discussed in Sect. 4.1.2 therein).\\
--- Royer et al. (2014) determined the abundances of 14 chemical species
for a homogeneous sample of 47 A0--A1 stars of comparatively low $v_{\rm e}\sin i$ 
($< 65$~km~s$^{-1}$). 
Comparing our [C/H] and [O/H] values with their results for 9 stars in common
(cf. their Table~4), we can confirm a reasonable agreement within the error bars;
i.e., the mean differences (ours$-$theirs) are $\langle\Delta$[C/H]$\rangle = -0.08$
(with the standard deviation of $\sigma = 0.08$) and  
$\langle\Delta$[O/H]$\rangle = -0.02$ ($\sigma = 0.11$).
According to their Fig.~10 (abundance range seen from the box 
size corresponding to the width of distribution), the results for [C/H] are  
$\sim -0.2$ to $-0.5$ (normal stars) and $\sim -0.4$ to $-0.7$ (CP stars) 
and those for [O/H] are $\sim 0.0$ to $-0.2$ (normal stars) and 
$\sim -0.2$ to $-0.4$ (CP stars). We may state that these values are 
roughly consistent with our results for stars of $T_{\rm eff} \sim$~9000--11000~K
(see figures~8e--8g).

\section{Conclusion}

Despite that many studies have been published regarding the photospheric chemical 
abundances in normal and chemically-peculiar A-type stars on the upper main sequence, 
only a limited number of spectroscopic investigations have been carried out so far 
concerning their abundances of CNO (light elements of astrophysical importance), 
which are known to be generally deficient in CP stars in contrast to many 
other heavier elements tending to be overabundant.

Motivated by this situation, we conducted a comprehensive spectroscopic 
study on the abundances of C, N, and O for 100 main-sequence 
stars of mostly A-type (late B through early F at 11000~K~$\gtsim T_{\rm eff} \gtsim$~7000~K) 
comprising normal stars as well as non-magnetic CP stars (Am and HgMn stars) 
in the projected rotational velocity range of 
0~km~s$^{-1}$~$\ltsim v_{\rm e}\sin i \ltsim$~100~km~s$^{-1}$,
based on the high-dispersion spectra obtained at Okayama Astrophysical Observatory 
(new observation for 29 targets) and Bohyunsan Astronomical Observatory.

Our aim was to investigate the abundance anomalies of CNO from qualitative as well as 
quantitative point of view, especially in terms of their mutual correlation 
or correlation with Fe, dependence upon stellar parameters ($T_{\rm eff}$, $v_{\rm e}\sin i$), 
and difference between normal and CP stars.

Regarding the method of analysis, we applied the spectrum-fitting technique to  
C~{\sc i}~5380, N~{\sc i}~7486, and O~{\sc i}~6156--8 lines and evaluated their equivalent 
widths, from which the non-LTE abundances, non-LTE corrections, and sensitivities 
to perturbations in atmospheric parameters were derived.   

The results of our analysis revealed the following observational characteristics 
regarding the CNO abundances of our sample stars:
\begin{itemize}
\item
C, N, and O are underabundant for almost all cases (irrespective of a star is classified 
as peculiar or normal, though with a tendency of larger anomaly for the former) 
typically in the range of $-1 \ltsim$~[C,N,O/H]~$\ltsim 0$), 
in contrast to [Fe/H] distributing around [Fe/H]~$\sim$~0.
\item
Moreover, distinctly large deficiencies as much as $\sim 2$~dex are shown for C or N 
by some CP stars ([C/H] for late Am stars or [N/H] for HgMn stars of late B-type)
\item
The inequality relation
$|[{\rm C}/{\rm H}]| > |[{\rm N}/{\rm H}]| > |[{\rm O}/{\rm H}]|$ 
appears to roughly hold regarding the typical extents of anomaly (deficiency),
which is consistent with the prediction from the recent model of atomic diffusion.  
\item
We confirmed that [C/H], [N/H], and [O/H] are anti-correlated with [Fe/H], which means
that the sense of chemical anomaly acts oppositely for CNO and heavier metals.
\item
The extent of CNO abundance peculiarity (deficiency) tends to be larger for lower 
$v_{\rm e}\sin i$, which becomes especially manifest when we pay attention only 
to 16 Hyades stars of the same primordial composition.
This may be in favor of the atomic diffusion theory for the cause of chemical anomaly, 
which would not work in the existence of efficient mixing by rapid rotation.  
\item
In addition, the dispersion of [C/H], [N/H], and [O/H] tends to grow (with 
the lower envelope of the distribution shifting toward lower values) with a decrease 
in $T_{\rm eff}$, which is consistent with recent diffusion model predicting that
the extent of CNO deficiency increases with decreasing $T_{\rm eff}$.
\end{itemize}

\bigskip
This research has made use of the SIMBAD database, operated at CDS, Strasbourg, France.
Data reduction was in part carried out by using the common-use data analysis 
computer system at the Astronomy Data Center (ADC) of the National Astronomical 
Observatory of Japan.


\newpage
\onecolumn

\setcounter{figure}{0}
\begin{figure}
\begin{center}
  \FigureFile(80mm,80mm){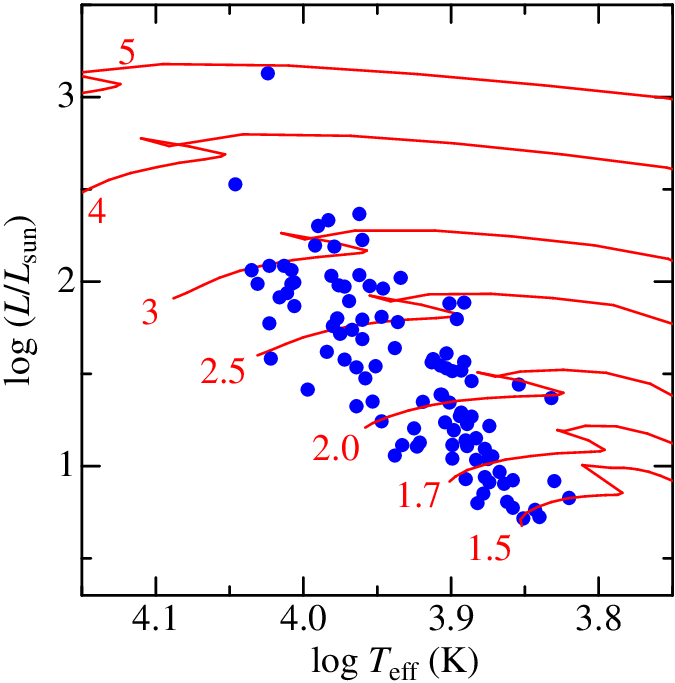}
\end{center}
\caption{
Our program stars plotted on the theoretical HR diagram ($\log (L/L_{\odot})$ 
vs. $\log T_{\rm eff}$), where $T_{\rm eff}$ was derived from colors (cf. section~3) 
and $L$ was evaluated from visual magnitude (corrected for interstellar extinction; 
Arenou, Grenon, \& G\'{o}mez 1992), Hipparcos parallax (van Leeuwen 2007), and 
bolometric correction (Flower 1996). Theoretical solar-metallicity tracks for 
7 different masses (1.5, 1.7, 2, 2.5, 3, 4, and 5~$M_{\odot}$), which were computed 
by Lejeune and Schaerer (2001), are also depicted by solid lines for comparison.
}
\label{fig:1}
\end{figure}

\setcounter{figure}{1}
\begin{figure}
  \begin{center}
    \FigureFile(150mm,190mm){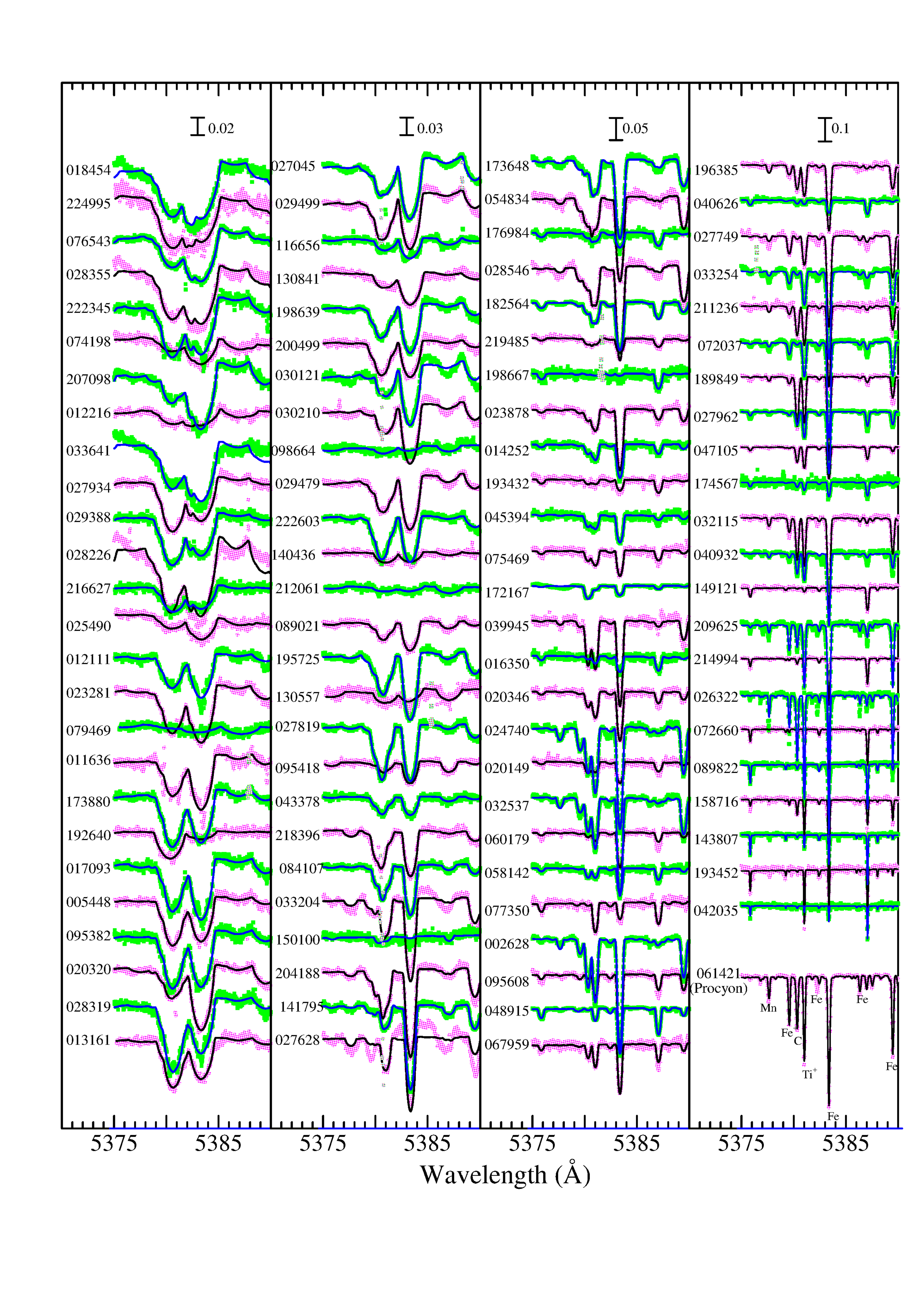}
  \end{center}
\caption{
Synthetic spectrum fitting in the 5375--5390~\AA\ region 
comprising the C~{\sc i}~5380 line. 
The best-fit theoretical spectra are shown by solid lines. 
The observed data are plotted by symbols, where those used 
in the fitting are colored in pink or green, while those rejected 
in the fitting (e.g., due to spectrum defect) are depicted in gray.
In each panel, the spectra are arranged in the descending order 
of $v_{\rm e} \sin i$ as in table~1, and vertical offsets of 0.04,
0.06, 0.08, and 0.15 (from the left to the right panels) are 
applied to each spectrum (indicated by the HD number) relative to 
the adjacent one. The case of Procyon (standard star) is 
displayed at the bottom of the rightmost panel.
}
\end{figure}

\setcounter{figure}{2}
\begin{figure}
  \begin{center}
    \FigureFile(150mm,190mm){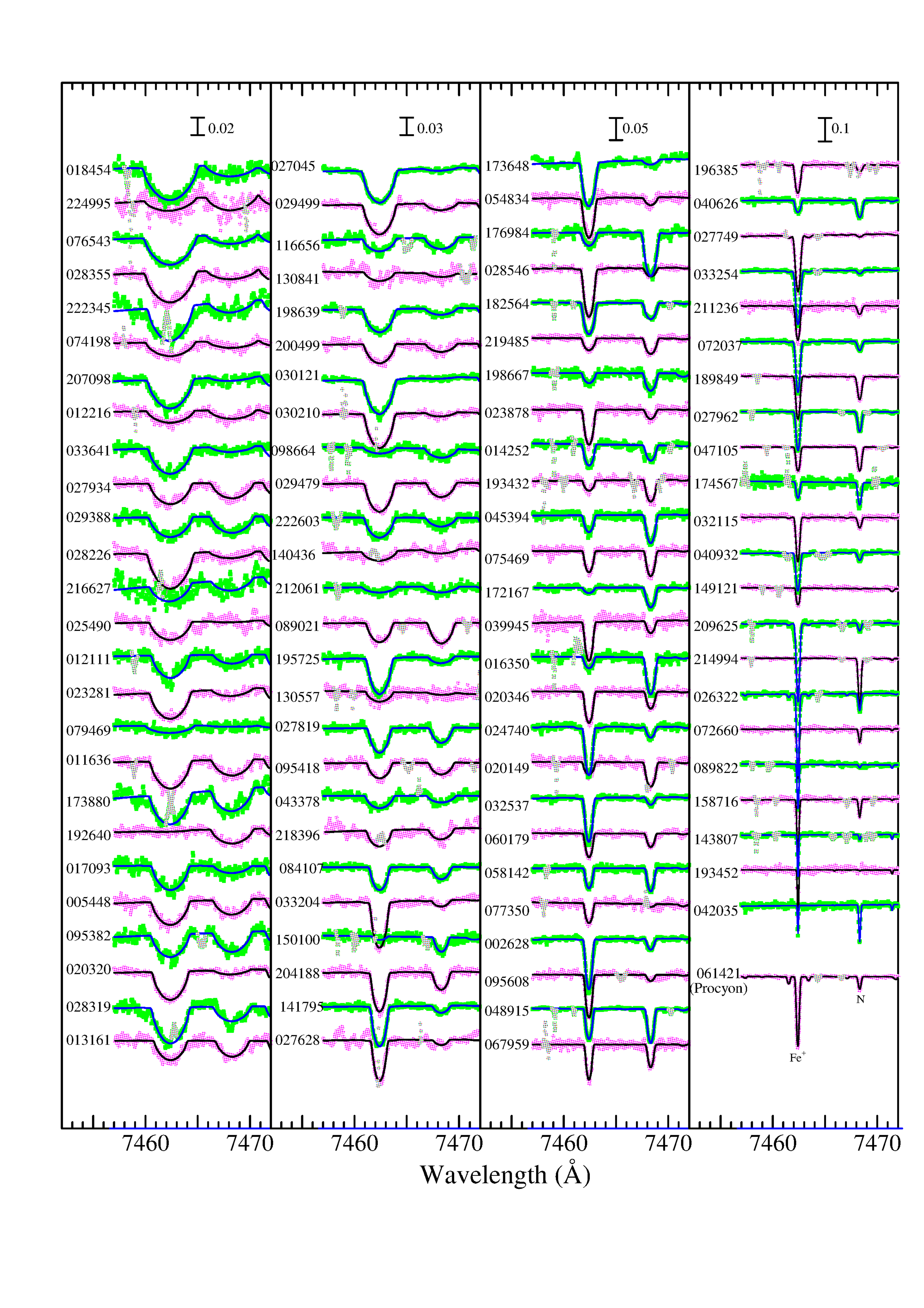}
  \end{center}
\caption{
Synthetic spectrum fitting in the 7457--7472~\AA\ region 
comprising the N~{\sc i}~7468 line.  Otherwise, the same as in figure~2.
}
\end{figure}

\setcounter{figure}{3}
\begin{figure}
  \begin{center}
    \FigureFile(150mm,190mm){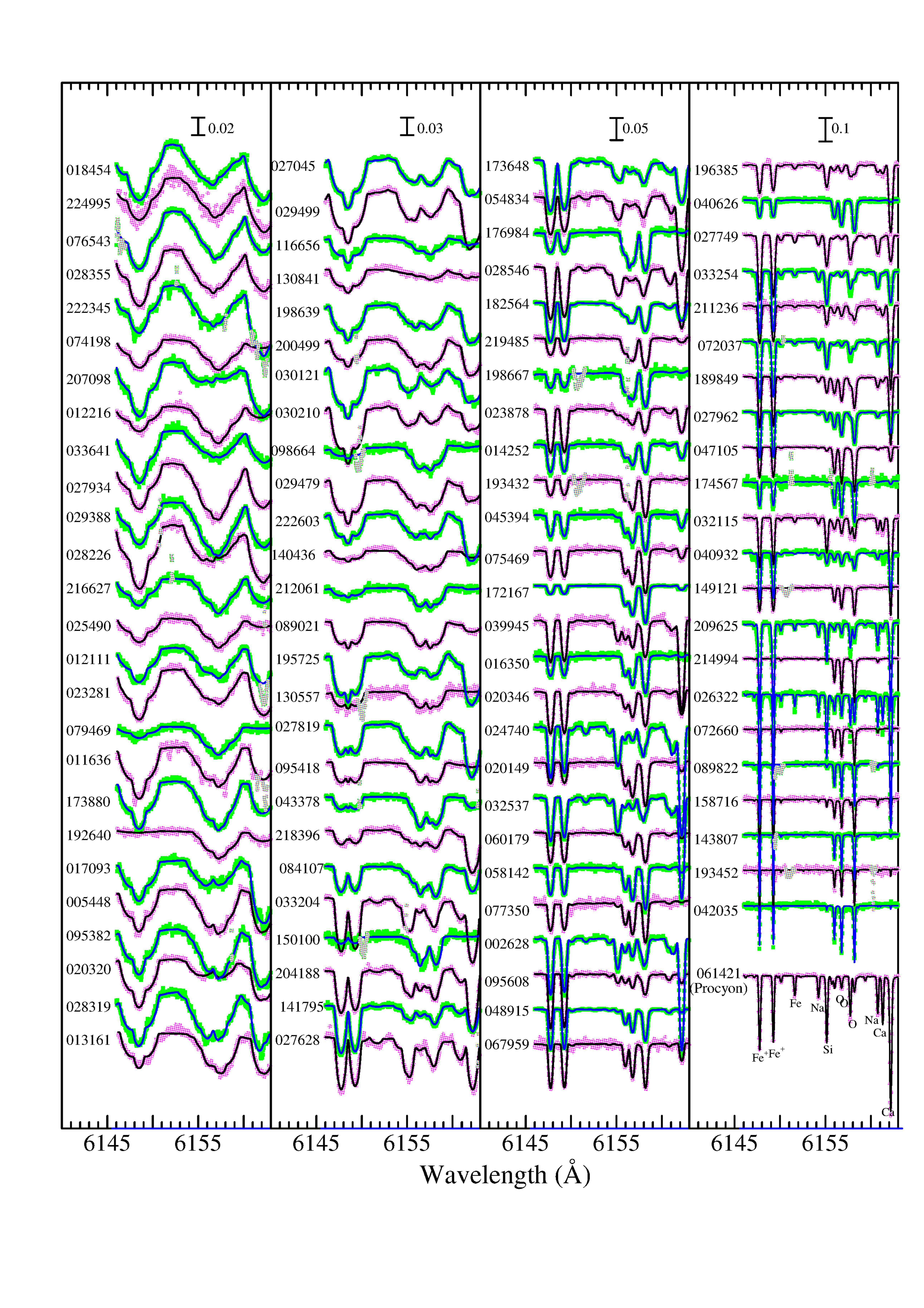}
  \end{center}
\caption{
Synthetic spectrum fitting in the 6146--6163~\AA\ region 
comprising the O~{\sc i}~6156--8 line.  Otherwise, the same as in figure~2.
}
\end{figure}

\setcounter{figure}{4}
\begin{figure}
  \begin{center}
    \FigureFile(110mm,150mm){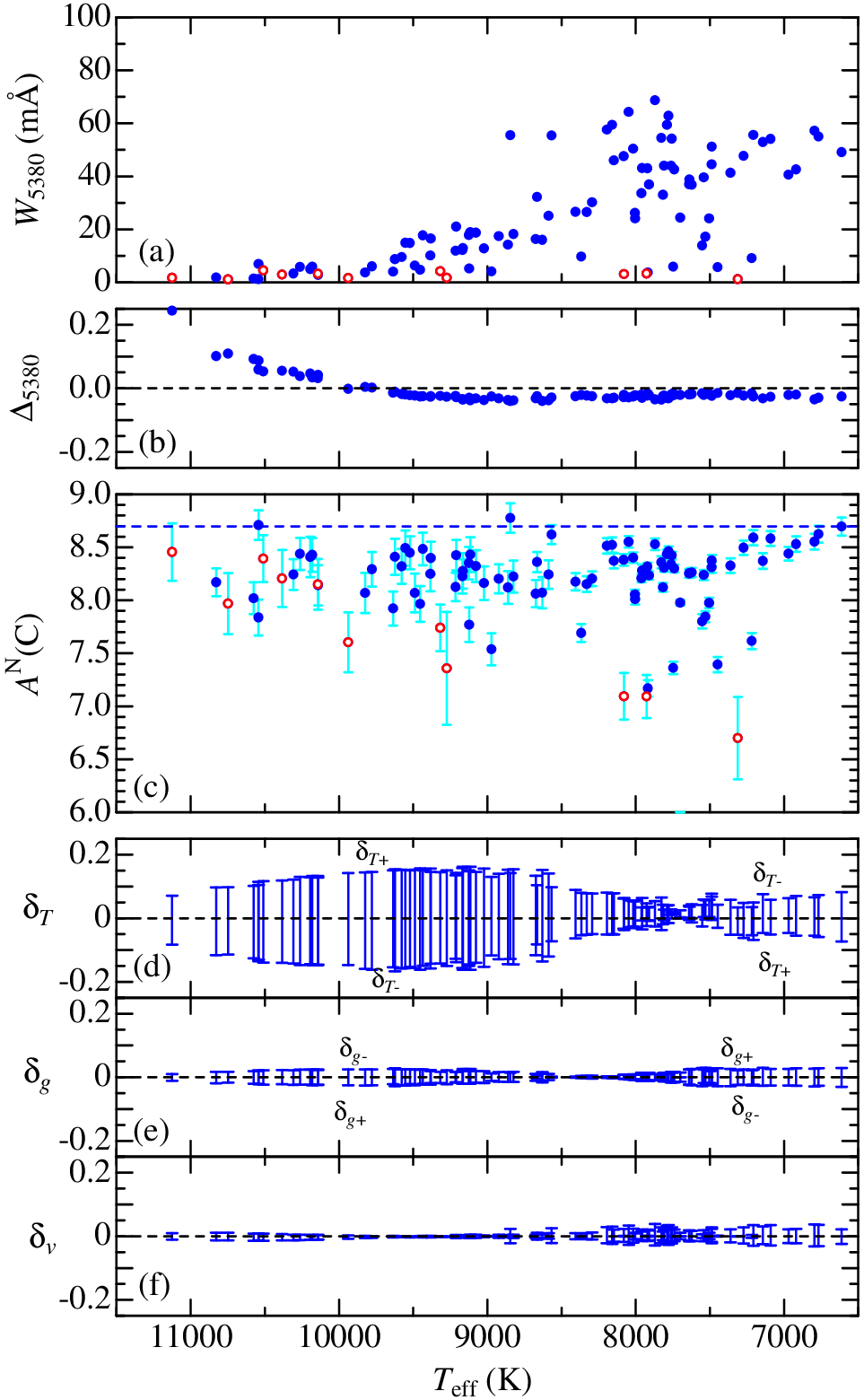}
  \end{center}
\caption{
Carbon abundance and C~{\sc i}~5380-related quantities 
plotted against $T_{\rm eff}$. 
(a) $W_{5380}$ (equivalent width of C~{\sc i} 5380), 
(b) $\Delta_{5380}$ (non-LTE correction for C~{\sc i} 5380),
(c) $A^{\rm N}$(C) (non-LTE abundance derived from C~{\sc i} 5380)
where the error bar denotes $\delta_{TgvW}$ (cf. subsection 4.3),
(d) $\delta_{T+}$ and $\delta_{T-}$ (abundance variations 
in response to $T_{\rm eff}$ changes of +3\% and $-$3\%), 
(e) $\delta_{g+}$ and $\delta_{g-}$ (abundance variations 
in response to $\log g$ changes by $+0.1$~dex and $-0.1$~dex), 
and (f) $\delta_{v+}$ and $\delta_{v-}$ (abundance 
variations in response to perturbing the $v_{\rm t}$ value
by +30\% and $-$30\%).
The data shown in open circles in panels (a) and (c) denote uncertain
results (which had better be regarded rather as upper limits) 
because of the weakness of lines (cf. subsection~4.3).
The abundance of Procyon, which is adopted as the reference, is indicated 
by the horizontal dashed line in panel (c). 
}
\end{figure}

\setcounter{figure}{5}
\begin{figure}
  \begin{center}
    \FigureFile(110mm,150mm){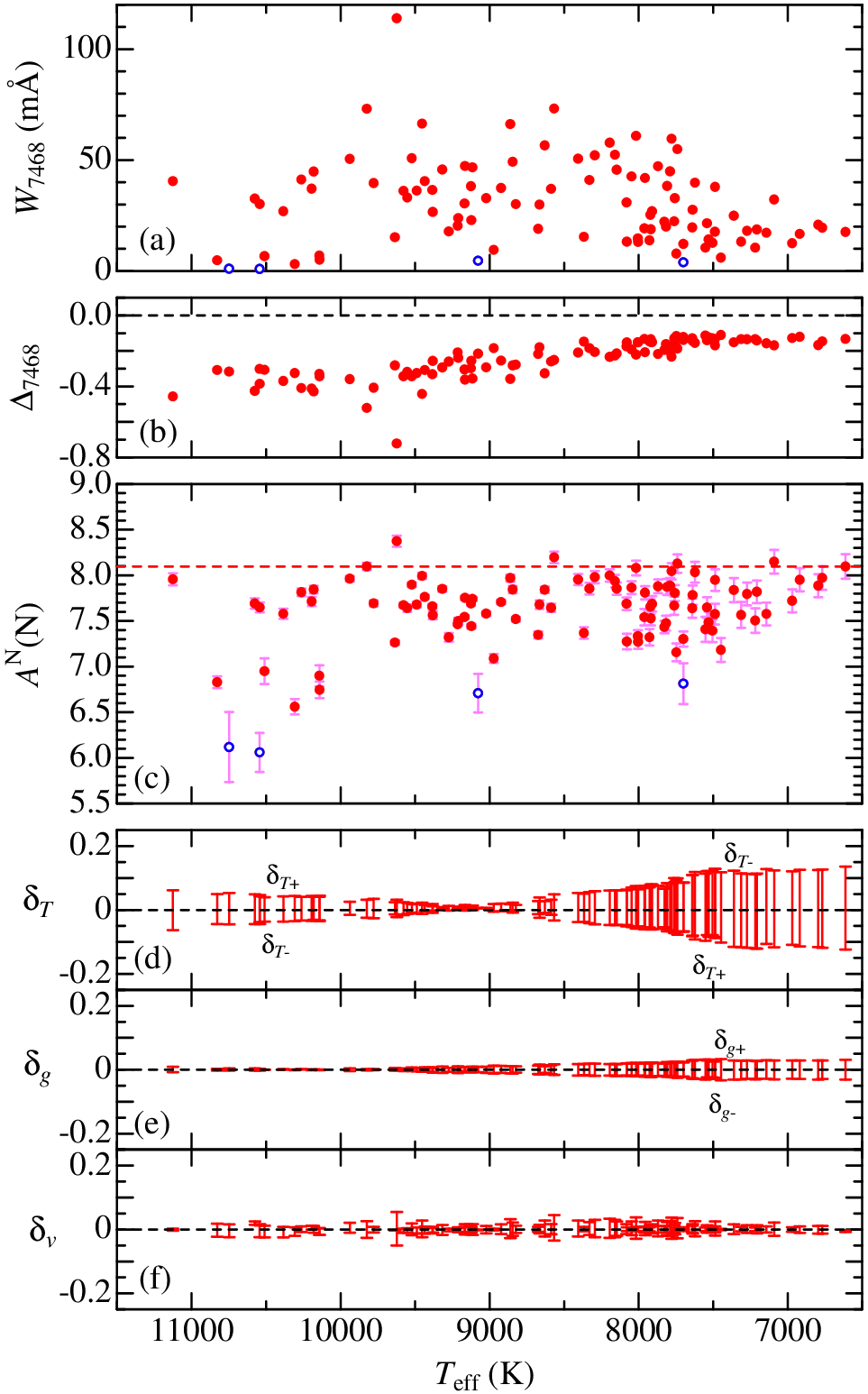}
  \end{center}
\caption{
Nitrogen abundance and N~{\sc i}~7486-related quantities 
plotted against $T_{\rm eff}$. 
Otherwise, the same as in figure 5. 
}
\end{figure}

\setcounter{figure}{6}
\begin{figure}
  \begin{center}
    \FigureFile(110mm,150mm){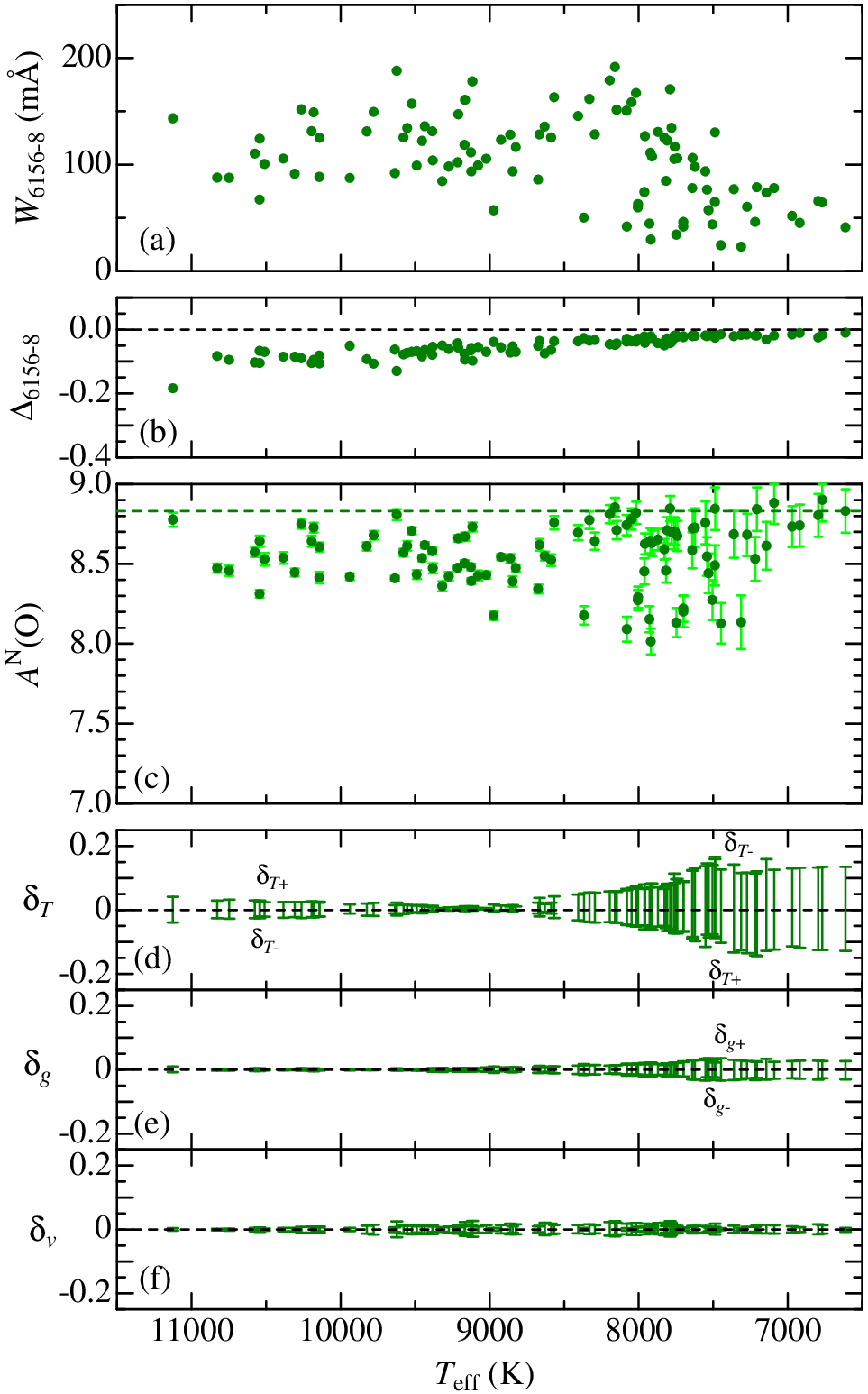}
  \end{center}
\caption{
Oxygen abundance and O~{\sc i}~6156--8-related quantities 
plotted against $T_{\rm eff}$. 
Otherwise, the same as in figure 5. 
}
\end{figure}

\setcounter{figure}{7}
\begin{figure}
  \begin{center}
    \FigureFile(160mm,200mm){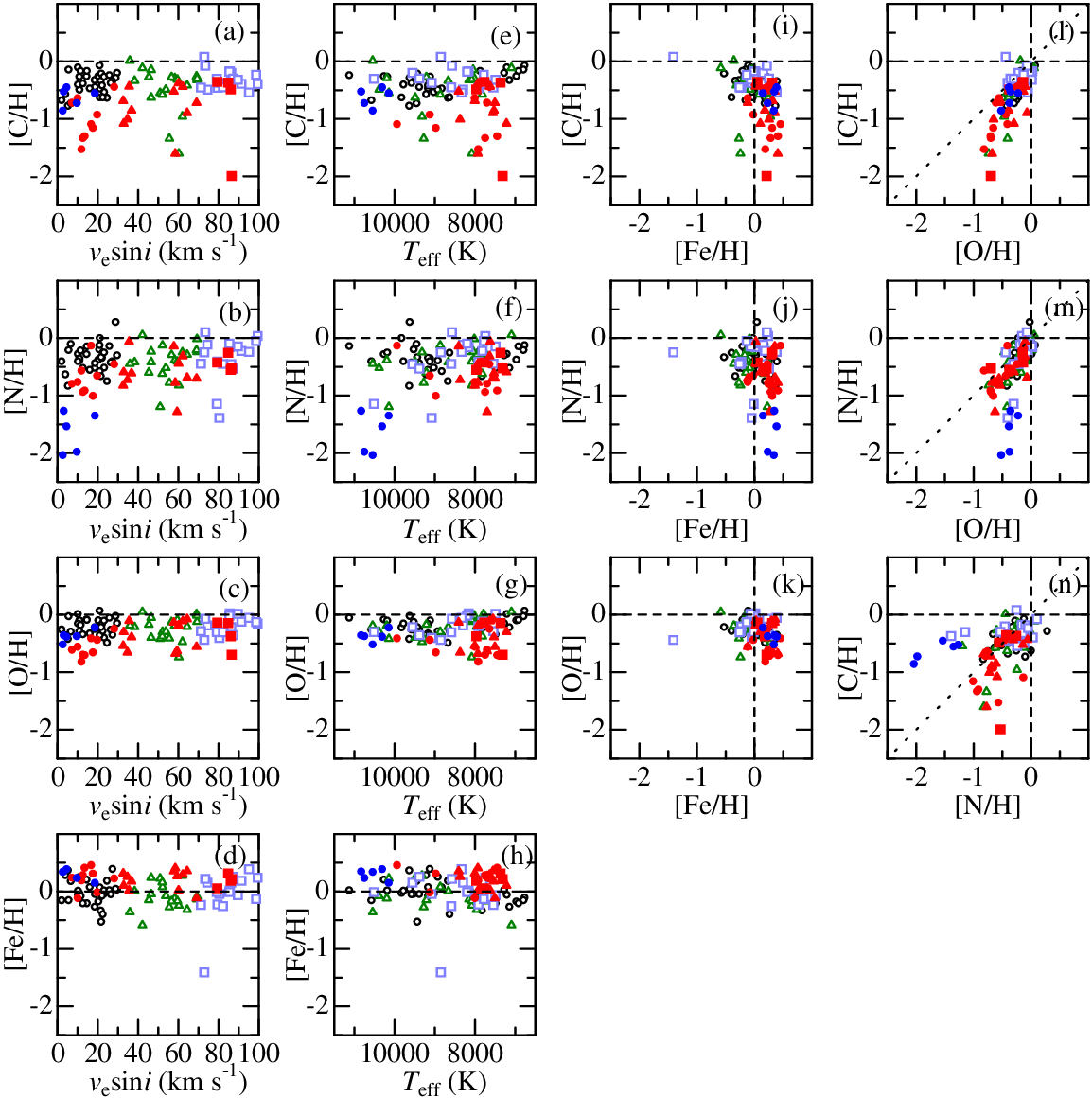}
  \end{center}
\caption{
Graphical display of how the abundances of C, N, and O (relative to Procyon) 
derived for 100 program stars depend upon $v_{\rm e}\sin i$ or $T_{\rm eff}$ 
and how they are mutually related with each other. 
In the 8 panels on the left-hand side are plotted [C/H], [N/H], [O/H], and [Fe/H] 
against $v_{\rm e}\sin i$ (panels a--d) and $T_{\rm eff}$ (panels e--h), 
while the 6 panels on the right-hand side show the correlation plot
for any combination between [C/H], [N/H], [O/H], and [Fe/H] (panels i--n). 
Stars of different $v_{\rm e}\sin i$ classes are discriminated by the types of symbols:
circles ($0 < v_{\rm e}\sin i < 30$~km~s$^{-1}$), 
triangles ($30 \le v_{\rm e}\sin i < 70$~km~s$^{-1}$), 
and squares ($70 \le v_{\rm e}\sin i < 100$~km~s$^{-1}$).
Those stars classified as chemically peculiar are highlighted by filled symbols:
red-filled symbols for Am stars and blue-filled ones for HgMn stars.
}
\end{figure}

\setcounter{figure}{8}
\begin{figure}
  \begin{center}
    \FigureFile(160mm,200mm){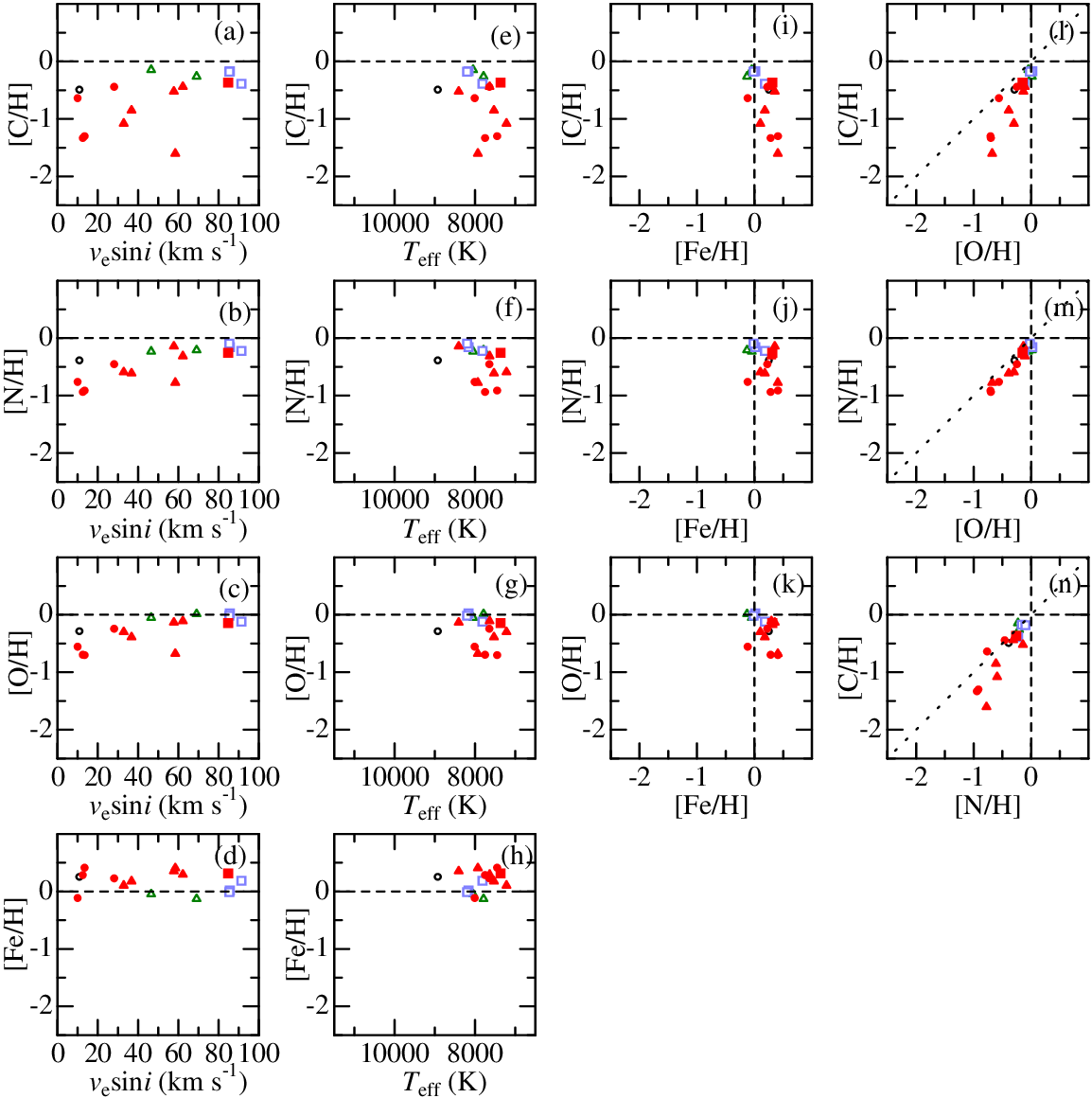}
  \end{center}
\caption{
Graphical display of how the abundances of C, N, and O (relative to Procyon) 
derived for 16 Hyades stars depend upon $v_{\rm e}\sin i$ or $T_{\rm eff}$ 
and how they are mutually related with each other. 
Otherwise, the same as in figure~8.
}
\end{figure}

\setcounter{table}{0}
\scriptsize
\renewcommand{\arraystretch}{0.8}
\setlength{\tabcolsep}{3pt}
\begin{longtable}{ccccccccccccl}
\caption{Stellar parameters and the resulting abundances of the program stars.}
\hline\hline
HD\# &  Name & Sp.Type & $T_{\rm eff}$ & $\log g$ & $v_{\rm t}$ & $v_{\rm e}\sin i$ & 
[Fe/H] & [C/H] & [N/H] & [O/H] & Group & Remark \\
(1) & (2) & (3) & (4) & (5) & (6) & (7) & (8) & (9) & (10) & (11) & (12) & (13)\\
\hline
\endhead
\hline
\endfoot
\hline
\multicolumn{13}{l}{\hbox to 0pt{\parbox{150mm}{\footnotesize
(1) HD number. (2) Bayer/Flamsteed name. (3) Spectral type taken from Hipparcos catalogue (ESA 1997).
(4) Effective temperature (in K). (5) Logarithm of surface gravity ($\log g$ in dex,
where $g$ is in unit of cm~s$^{-2}$). (6) Microturbulent velocity (in km~s$^{-1}$).  
(7) Projected rotational velocity (in km~s$^{-1}$) derived from 6146--6163~\AA\ fitting. 
(8)--(11) Abundances of Fe (from 6146--6163~\AA\ fitting), C, N, and O relative to the standard 
star Procyon ([X/H] $\equiv$ $A_{\rm X}$(star) $-$ $A_{\rm X}$(Procyon)), where the parenthesized
values denote uncertain results (which may as well be regarded as representing the upper limits).
The $A^{\rm N}_{\rm X}$(Procyon) itself is shown in the last row for Procyon, where $A$ is 
the logarithmic number abundance relative to H expressed in the usual normalization of $A_{\rm H} = 12$.  
(12) Group of the data source (cf. table~2).
(13) Specific remark [spectroscopic binary (SB) or radial velocity variable (V), 
chemical peculiarity type (Am or HgMn), membership of Hyades cluster (H); see section~2].
}}}
\endlastfoot
\hline
018454 &  4~Eri  & A5IV/V     &  7740 & 4.07 & 3.9 & 99.5 & +0.24 & $-$0.40 & +0.03 & $-$0.16 & B & V      \\
224995 & 31~Psc  & A6V        &  7779 & 3.64 & 4.0 & 98.7 & $-$0.13 & $-$0.23 & $-$0.05 & $-$0.13 & D & V      \\
076543 & $o^{1}$~Cnc  & A5III      &  8330 & 4.18 & 3.9 & 95.2 & +0.38 & $-$0.54 & $-$0.24 & $-$0.06 & B & SB       \\
028355 & 79~Tau  & A7V        &  7809 & 3.98 & 4.0 & 91.5 & +0.19 & $-$0.39 & $-$0.22 & $-$0.12 & B & V?,Hyades \\
222345 & $\omega^{1}$~Aqr & A7IV       &  7487 & 3.88 & 3.8 & 89.7 & $-$0.07 & $-$0.32 & $-$0.15 & +0.01 & B & SB      \\
074198 & $\gamma$~Cnc  & A1IV       &  9381 & 4.11 & 2.8 & 87.4 & +0.25 & $-$0.30 & $-$0.53 & $-$0.36 & B & SB      \\
207098 & $\delta$~Cap  & A5mF2 (IV) &  7312 & 4.06 & 3.6 & 86.6 & +0.21 & ($-$2.00) & $-$0.53 & $-$0.70 & B & SB, Am    \\
012216 & 50~Cas  & A2V        &  9553 & 3.90 & 2.6 & 86.4 & +0.15 & $-$0.20 & $-$0.46 & $-$0.22 & B &  SB2     \\
033641 & $\mu$~Aur  & A4m        &  7961 & 4.21 & 4.0 & 86.2 & +0.18 & $-$0.49 & $-$0.55 & $-$0.38 & B & V, Am    \\
027934 & $\kappa^{1}$~Tau  & A7IV-V     &  8159 & 3.84 & 4.0 & 85.7 & +0.02 & $-$0.17 & $-$0.16 & +0.02 & B & SB?, Hyades \\
029388 & 90~Tau  & A6V        &  8194 & 3.88 & 4.0 & 85.5 & $-$0.01 & $-$0.18 & $-$0.10 & $-$0.02 & B & SB1, Hyades \\
028226 &            & Am         &  7361 & 4.01 & 3.6 & 84.9 & +0.31 & $-$0.37 & $-$0.26 & $-$0.15 & B & SB2, Am, Hyades \\
216627 & $\delta$~Aqr  & A3V        &  8587 & 3.59 & 3.7 & 82.3 & $-$0.25 & $-$0.45 & $-$0.45 & $-$0.31 & B & V      \\
025490 & $\nu$~Tau  & A1V        &  9077 & 3.93 & 3.2 & 80.5 & $-$0.05 & $-$0.37 & ($-$1.39) & $-$0.41 & B &       \\
012111 & 48~Cas  & A3IV       &  7910 & 4.08 & 4.0 & 79.6 & $-$0.23 & $-$0.46 & $-$0.41 & $-$0.20 & B &  SB     \\
023281 &            & A5m        &  7761 & 4.19 & 4.0 & 79.4 & +0.05 & $-$0.36 & $-$0.43 & $-$0.14 & B & Am    \\
079469 & $\theta$~Hya  & B9.5V      & 10510 & 4.20 & 1.4 & 79.2 & $-$0.02 & ($-$0.30) & $-$1.15 & $-$0.30 & B & SB      \\
011636 &  $\beta$~Ari  & A5V...     &  8294 & 4.12 & 3.9 & 74.5 & +0.15 & $-$0.49 & $-$0.12 & $-$0.19 & B &  SB     \\
173880 & 111~Her & A5III      &  8567 & 4.27 & 3.8 & 73.5 & +0.22 & $-$0.08 & +0.10 & $-$0.07 & B & SB?      \\
192640 & 29~Cyg  & A2V        &  8845 & 3.86 & 3.5 & 73.0 & $-$1.41 & +0.08 & $-$0.25 & $-$0.44 & B & V      \\
017093 & 38~Ari  & A7III-IV   &  7541 & 3.95 & 3.8 & 71.3 & $-$0.23 & $-$0.46 & $-$0.45 & $-$0.29 & B & V       \\
095382 & 59~Leo  & A5III      &  8017 & 3.95 & 4.0 & 69.3 & $-$0.09 & $-$0.29 & $-$0.02 & $-$0.01 & B &       \\
005448 & $\mu$~And  & A5V        &  8147 & 3.82 & 4.0 & 69.3 & $-$0.14 & $-$0.32 & $-$0.24 & $-$0.12 & B &       \\
028319 & $\theta^{2}$~Tau  & A7III      &  7789 & 3.68 & 4.0 & 69.1 & $-$0.13 & $-$0.26 & $-$0.21 & +0.01 & B & SB1, Hyades \\
020320 & $\zeta$~Eri  & A5m        &  7505 & 3.91 & 3.8 & 69.1 & $-$0.12 & $-$0.72 & $-$0.70 & $-$0.56 & B & SB, Am    \\
027045 & $\omega^{2}$~Tau  & A3m        &  7552 & 4.26 & 3.8 & 64.4 & +0.36 & $-$0.89 & $-$0.69 & $-$0.07 & B & SB, Am    \\
013161 & $\beta$~Tri  & A5III      &  7957 & 3.68 & 4.0 & 64.4 & $-$0.32 & $-$0.42 & $-$0.29 & $-$0.21 & B &  SB2     \\
029499 &            & A5m        &  7638 & 4.08 & 3.9 & 62.3 & +0.29 & $-$0.44 & $-$0.31 & $-$0.11 & B & V, Am, Hyades \\
116656 & $\zeta$~UMa  & A2V        &  9317 & 4.10 & 2.9 & 62.1 & +0.28 & ($-$0.96) & $-$0.25 & $-$0.47 & B & SB2       \\
130841 & $\alpha^{2}$~Lib  & A3IV       &  8079 & 3.96 & 4.0 & 60.3 & $-$0.24 & ($-$1.60) & $-$0.82 & $-$0.74 & B &  SB     \\
198639 & 56~Cyg  & A4me...    &  7921 & 4.09 & 4.0 & 60.2 & +0.02 & $-$0.38 & $-$0.44 & $-$0.19 & B & V?, Am    \\
200499 & $\eta$~Cap  & A5V        &  8081 & 3.95 & 4.0 & 59.6 & $-$0.17 & $-$0.31 & $-$0.41 & $-$0.09 & B & V      \\
030121 &  4~Cam  & A3m        &  7700 & 3.98 & 3.9 & 59.4 & +0.27 & $\cdots$ & ($-$1.28) & $-$0.63 & B & Am    \\
030210 &            & Am...      &  7927 & 3.94 & 4.0 & 58.5 & +0.40 & ($-$1.60) & $-$0.77 & $-$0.68 & B & SB1?, Am, Hyades \\
098664 & $\sigma$~Leo  & B9.5Vs     & 10194 & 3.75 & 1.8 & 58.3 & $-$0.12 & $-$0.29 & $-$0.38 & $-$0.19 & C & SB      \\
029479 & $\sigma^{1}$~Tau  & A4m        &  8406 & 4.14 & 3.9 & 57.8 & +0.35 & $-$0.52 & $-$0.14 & $-$0.14 & B & SB, Am, Hyades \\
222603 & $\lambda$~Psc  & A7V        &  7757 & 3.99 & 4.0 & 56.6 & $-$0.17 & $-$0.27 & $-$0.29 & $-$0.13 & B & SB      \\
140436 & $\gamma$~CrB  & A1Vs       &  9274 & 3.89 & 3.0 & 55.6 & $-$0.27 & ($-$1.34) & $-$0.78 & $-$0.41 & B &       \\
212061 & $\gamma$~Aqr  & A0V        & 10384 & 3.95 & 1.5 & 53.8 & $-$0.08 & ($-$0.49) & $-$0.52 & $-$0.29 & B & SB     \\
089021 & $\lambda$~UMa  & A2IV       &  8861 & 3.61 & 3.5 & 52.2 & +0.08 & $-$0.57 & $-$0.13 & $-$0.30 & B & V     \\
195725 & $\theta$~Cep  & A7III      &  7816 & 3.74 & 4.0 & 52.1 & +0.16 & $-$0.57 & $-$0.62 & $-$0.37 & B & SB2      \\
130557 &            & B9Vsvar... & 10142 & 3.85 & 1.8 & 51.0 & +0.23 & ($-$0.55) & $-$1.20 & $-$0.42 & C &       \\
095418 & $\beta$~UMa  & A1V        &  9489 & 3.85 & 2.7 & 46.5 & +0.24 & $-$0.63 & $-$0.42 & $-$0.40 & B & SB      \\
027819 & $\delta^{2}$~Tau  & A7V        &  8047 & 3.95 & 4.0 & 46.5 & $-$0.05 & $-$0.15 & $-$0.23 & $-$0.05 & B &  SB, Hyades \\
043378 &  2~Lyn  & A2Vs       &  9210 & 4.09 & 3.0 & 45.3 & $-$0.15 & $-$0.27 & $-$0.60 & $-$0.17 & B &  V?     \\
218396 &            & A5V        &  7091 & 4.06 & 3.3 & 42.2 & $-$0.59 & $-$0.11 & +0.05 & +0.05 & B &       \\
084107 & 15~Leo  & A2IV       &  8665 & 4.31 & 3.7 & 38.3 & +0.01 & $-$0.34 & $-$0.42 & $-$0.21 & B &       \\
033204 &            & A5m        &  7530 & 4.06 & 3.8 & 36.8 & +0.18 & $-$0.85 & $-$0.61 & $-$0.39 & B & Am, Hyades \\
150100 & 16~Dra  & B9.5Vn     & 10542 & 3.84 & 1.4 & 35.9 & $-$0.36 & +0.01 & $-$0.45 & $-$0.19 & C & V     \\
204188 &            & A8m        &  7622 & 4.21 & 3.9 & 35.6 & +0.02 & $-$0.43 & $-$0.06 & $-$0.11 & B & SB, Am    \\
141795 & $\epsilon$~Ser  & A2m        &  8367 & 4.24 & 3.9 & 34.8 & +0.25 & $-$1.01 & $-$0.73 & $-$0.65 & B & V, Am    \\
027628 & 60~Tau  & A3m        &  7218 & 4.05 & 3.5 & 32.9 & +0.10 & $-$1.08 & $-$0.59 & $-$0.30 & B & SB1, Am, Hyades \\
173648 & $\zeta^{1}$~Lyr  & Am         &  8004 & 3.90 & 4.0 & 32.6 & +0.32 & $-$0.69 & $-$0.82 & $-$0.54 & B & SB1, Am    \\
054834 &            & A9V        &  7273 & 4.21 & 3.5 & 29.6 & +0.03 & $-$0.20 & $-$0.30 & $-$0.15 & D &       \\
176984 & 14~Aql  & A1V        &  9623 & 3.42 & 2.5 & 28.9 & +0.04 & $-$0.29 & +0.28 & $-$0.02 & D &  V?     \\
028546 & 81~Tau  & Am         &  7640 & 4.17 & 3.9 & 28.2 & +0.23 & $-$0.44 & $-$0.46 & $-$0.25 & B & V?, Am, Hyades \\
182564 & $\pi$~Dra  & A2IIIs     &  9125 & 3.80 & 3.1 & 27.3 & +0.39 & $-$0.35 & $-$0.41 & $-$0.35 & A &       \\
219485 &            & A0V        &  9577 & 3.81 & 2.5 & 26.5 & $-$0.05 & $-$0.38 & $-$0.42 & $-$0.26 & D &       \\
198667 &  5~Aqr  & B9III      & 11125 & 3.42 & 0.9 & 25.8 & +0.02 & ($-$0.24) & $-$0.14 & $-$0.05 & C & V      \\
023878 & $\tau^{7}$~Eri  & A1V        &  8674 & 3.80 & 3.7 & 24.6 & +0.18 & $-$0.63 & $-$0.75 & $-$0.49 & D & V?      \\
014252 & 10~Tri  & A2V        &  9023 & 3.64 & 3.3 & 23.2 & $-$0.04 & $-$0.53 & $-$0.52 & $-$0.40 & D & V      \\
193432 &  $\nu$~Cap  & B9IV       & 10180 & 3.91 & 1.8 & 23.1 & +0.02 & $-$0.27 & $-$0.25 & $-$0.10 & D & V?       \\
045394 & 16~Gem  & A2Vs       &  8630 & 3.42 & 3.7 & 22.5 & $-$0.40 & $-$0.62 & $-$0.25 & $-$0.28 & D &       \\
075469 &            & A2Vs       &  9165 & 3.51 & 3.1 & 22.1 & $-$0.08 & $-$0.42 & $-$0.34 & $-$0.16 & D &       \\
172167 &  $\alpha$~Lyr  & A0Vvar     &  9435 & 3.99 & 2.7 & 21.7 & $-$0.53 & $-$0.21 & $-$0.33 & $-$0.21 & A & V      \\
039945 &            & A5V        &  7827 & 3.36 & 4.0 & 21.5 & $-$0.33 & $-$0.34 & $-$0.66 & $-$0.24 & D &       \\
020346 &            & A2IV       &  8824 & 3.56 & 3.5 & 21.0 & +0.07 & $-$0.47 & $-$0.58 & $-$0.36 & D & SB?      \\
016350 &            & B9.5V      &  9824 & 3.72 & 2.2 & 21.0 & $-$0.03 & $-$0.63 & +0.00 & $-$0.22 & D &       \\
024740 & 32~Tau  & F2IV       &  6768 & 3.77 & 2.7 & 20.9 & $-$0.11 & $-$0.07 & $-$0.12 & +0.07 & D & V      \\
020149 &            & A1Vs       &  9522 & 3.99 & 2.6 & 20.7 & $-$0.05 & $-$0.25 & $-$0.20 & $-$0.12 & D & SB?      \\
032537 &  9~Aur  & F0V        &  6970 & 4.07 & 3.1 & 20.3 & $-$0.18 & $-$0.26 & $-$0.38 & $-$0.10 & D & SB      \\
060179 & $\alpha$~Gem  & A2Vm       &  9122 & 3.88 & 3.2 & 19.7 & $-$0.02 & $-$0.93 & $-$0.65 & $-$0.44 & B & SB1, Am    \\
077350 & $\nu$~Cnc  & A0III      & 10141 & 3.68 & 1.8 & 18.6 & +0.15 & $-$0.56 & $-$1.35 & $-$0.23 & C & SB, HgMn    \\
058142 & 21~Lyn  & A1V        &  9384 & 3.74 & 2.8 & 18.6 & $-$0.05 & $-$0.45 & $-$0.44 & $-$0.25 & D & V      \\
002628 & 28~And  & A7III      &  7143 & 3.48 & 3.3 & 18.5 & $-$0.27 & $-$0.32 & $-$0.52 & $-$0.22 & D &       \\
095608 & 60~Leo  & A1m        &  8972 & 4.20 & 3.3 & 17.6 & +0.31 & $-$1.16 & $-$1.01 & $-$0.66 & B & Am    \\
048915 & $\alpha$~CMa  & A0m...     &  9938 & 4.31 & 2.1 & 16.7 & +0.45 & ($-$1.09) & $-$0.13 & $-$0.41 & A & SB, Am    \\
067959 &            & A1V        &  9168 & 3.65 & 3.1 & 16.0 & +0.07 & $-$0.47 & $-$0.55 & $-$0.33 & D &       \\
196385 &            & A9V        &  6919 & 4.23 & 3.0 & 14.6 & $-$0.21 & $-$0.17 & $-$0.15 & $-$0.09 & D &       \\
040626 &            & B9.5IV     & 10263 & 4.00 & 1.7 & 14.4 & +0.20 & $-$0.26 & $-$0.28 & $-$0.08 & D &       \\
027749 & 63~Tau  & A1m        &  7448 & 4.21 & 3.7 & 13.5 & +0.41 & $-$1.30 & $-$0.91 & $-$0.70 & B & SB1, Am, Hyades \\
211236 &            & A8/A9IV/V  &  7488 & 3.96 & 3.8 & 12.6 & $-$0.21 & $-$0.39 & $-$0.52 & $-$0.34 & D &       \\
033254 & 16~Ori  & A2m        &  7747 & 4.14 & 3.9 & 12.6 & +0.28 & $-$1.33 & $-$0.94 & $-$0.70 & B & SB, Am, Hyades \\
072037 &  2~UMa  & A2m        &  7918 & 4.16 & 4.0 & 11.9 & +0.19 & $-$1.53 & $-$0.57 & $-$0.82 & B & Am    \\
189849 & 15~Vul  & A4III      &  7870 & 3.62 & 4.0 & 11.5 & $-$0.08 & $-$0.17 & $-$0.22 & $-$0.18 & A & SB      \\
047105 & $\gamma$~Gem  & A0IV       &  9115 & 3.49 & 3.2 & 10.9 & $-$0.03 & $-$0.27 & $-$0.35 & $-$0.10 & A & SB      \\
027962 & $\delta^{3}$~Tau  & A2IV       &  8923 & 3.94 & 3.4 & 10.9 & +0.25 & $-$0.49 & $-$0.39 & $-$0.29 & A & SB, Hyades \\
174567 &            & A0Vs       &  9778 & 3.59 & 2.3 & 10.2 & +0.01 & $-$0.40 & $-$0.40 & $-$0.15 & D &       \\
032115 &            & A8IV       &  7207 & 4.13 & 3.4 & 10.1 & $-$0.06 & $-$0.11 & $-$0.28 & +0.01 & D & V      \\
040932 & $\mu$~Ori  & Am...      &  8005 & 3.93 & 4.0 & 10.0 & $-$0.12 & $-$0.64 & $-$0.76 & $-$0.56 & B & SB1, Am, Hyades \\
149121 & 28~Her  & B9.5III    & 10748 & 3.89 & 1.2 &  9.6 & +0.23 & ($-$0.73) & ($-$1.98) & $-$0.37 & C & HgMn    \\
209625 & 32~Aqr  & A5m        &  7700 & 3.87 & 3.9 &  7.2 & +0.24 & $-$0.72 & $-$0.79 & $-$0.61 & D & SB1, Am    \\
214994 & $o$~Peg  & A1IV       &  9453 & 3.64 & 2.7 &  6.6 & +0.18 & $-$0.73 & $-$0.10 & $-$0.30 & A & V     \\
026322 & 44~Tau  & F2IV-V     &  6795 & 3.46 & 2.8 &  5.5 & $-$0.15 & $-$0.14 & $-$0.21 & $-$0.03 & D &       \\
072660 &            & A1V        &  9635 & 3.97 & 2.5 &  5.2 & +0.37 & $-$0.77 & $-$0.83 & $-$0.42 & D &       \\
089822 &            & A0sp...    & 10307 & 3.89 & 1.6 &  4.5 & +0.39 & $-$0.45 & $-$1.54 & $-$0.39 & C & SB2, HgMn    \\
158716 &            & A1V        &  9214 & 4.30 & 3.0 &  3.9 & +0.28 & $-$0.57 & $-$0.63 & $-$0.36 & D &       \\
143807 & $\iota$~CrB  & A0p...     & 10828 & 4.06 & 1.1 &  3.1 & +0.35 & $-$0.53 & $-$1.27 & $-$0.36 & D & SB, HgMn    \\
193452 &            & B9.5III/IV & 10543 & 4.15 & 1.4 &  2.6 & +0.34 & $-$0.86 & ($-$2.04) & $-$0.52 & C & SB1, HgMn    \\
042035 &            & B9V        & 10575 & 3.82 & 1.4 &  2.0 & $-$0.16 & $-$0.68 & $-$0.40 & $-$0.26 & D & V      \\
\hline
061421 & Procyon & F5IV-V     &  6612 & 4.00 & 2.0 &  6.7 & 7.47 & 8.70 & 8.10 & 8.83 & E &  SB     \\
\end{longtable}

\setcounter{table}{1}
\begin{table}[h]
\small
\caption{Basic information of the adopted observational data.}
\begin{center}
\begin{tabular}{ccccccc}\hline\hline
Group & $^{\#}$Instr. & Obs.Time & Resolution & Number & Star Type & Reference \\
\hline
$^{\dagger}$A & HIDES & 2008 Oct & 100000 & 7 & sharp-line A & Takeda et al. (2012) \\
B & BOES & 2008 Jan/Sep, 2009 Jan & 45000 & 56 & sharp/broad-line A & Takeda et al. (2008, 2009) \\
C & HIDES & 2012 May  & 70000 & 8 & sharp-line late B & Takeda et al. (2014) \\
D & HIDES & 2017 Aug/Nov & 100000 & 29 & sharp-line early F and A & This study (cf. section~2)\\
$^{*}$E & HIDES & 2001 Feb & 70000 & 1 & Procyon & Takeda et al. (2005a) \\
\hline
\end{tabular}
\end{center}
$^{\dagger}$Only for HD~172167 (Vega), we adopted the OAO/HIDES spectra of high-S/N ($\sim 2000$) 
and high-resolution ($\sim 100000$) published by Takeda, Kawanomoto, and Ohishi (2007).\\
$^{*}$Regarding the Procyon spectra used for 7457--7472~\AA\ fitting, we used the data
published by Allende Prieto et al. (2004).\\
$^{\#}$HIDES and BOES denote ``HIgh Dispersion Echelle Spectrograph'' at Okayama Astrophysical 
Observatory and ``Bohyunsan Observatory Echelle Spectrograph'' at Bohyunsan Optical Astronomy 
Observatory, respectively.
\end{table}

\setcounter{table}{2}
\begin{table}[h]
\small
\caption{Outline of spectrum-fitting analysis in this study.}
\begin{center}
\begin{tabular}{ccccc}\hline\hline
Purpose & fitting range (\AA) & abundances varied$^{*}$ & atomic data source & figure \\
\hline
C abundance from C~{\sc i} 5380   & 5375--5390 & C, Ti, Fe & KB95m1 & figure~2 \\
N abundance from N~{\sc i} 7468  & 7457--7472 & N, Fe & KB95m2 & figure~3 \\
O abundance from O~{\sc i} 6156--8   & 6146--6163 & O, Na, Si, Ca, Fe & KB95 & figure~4 \\
\hline
\end{tabular}
\end{center}
$^{*}$ The abundances of all other elements than these were fixed in the fitting. \\
KB95m1 --- All the atomic line data presented in Kurucz and Bell (1995) were used, 
excepting that the contribution of Fe~{\sc i} 5382.474 ($\chi_{\rm low} = 4.371$~eV) 
was neglected (because we found its $gf$ value to be erroneously too large). \\
KB95m2 --- All the atomic line data were taken from Kurucz and Bell (1995), excepting
that the contribution of S~{\sc i} 7468.588 ($\chi_{\rm low} = 7.867$~eV) was neglected 
(because we found its $gf$ value to be erroneously too large). \\
KB95 --- All the atomic line data given in Kurucz and Bell (1995) were used unchanged.
\end{table}

\setcounter{table}{3}
\begin{table}[h]
\small
\caption{Adopted atomic data of relevant CNO lines.}
\begin{center}
\begin{tabular}{ccccccccc}\hline\hline
Line & Multiplet & Equivalent & $\lambda$ & $\chi_{\rm low}$ & $\log gf$ & Gammar & Gammas & Gammaw\\
     &  No.           &  Width  & (\AA) & (eV) & (dex) & (dex) & (dex) & (dex) \\  
\hline
C~{\sc i} 5380 & (11)  & $W_{5380}$ &  5380.337 & 7.685 & $-1.842$ & (7.89) & $-$4.66 & ($-$7.36)\\
\hline
N~{\sc i} 7468 & (3) & $W_{7468}$ & 7468.312 & 10.336 & $-0.270$  & 8.64 & $-$5.40 & ($-$7.60)\\
\hline
O~{\sc i} 6156--8 & (10) & $W_{6156-8}$ & 6155.961 & 10.740 & $-1.401$ & 7.60 & $-$3.96 & ($-$7.23)\\
(9 components)&  &  & 6155.971 & 10.740 & $-1.051$ & 7.61 & $-$3.96 & ($-$7.23)\\
         &         &          & 6155.989 & 10.740 & $-1.161$ & 7.61 & $-$3.96 & ($-$7.23)\\
         &         &          & 6156.737 & 10.740 & $-1.521$ & 7.61 & $-$3.96 & ($-$7.23)\\
         &         &          & 6156.755 & 10.740 & $-0.931$ & 7.61 & $-$3.96 & ($-$7.23)\\
         &         &          & 6156.778 & 10.740 & $-0.731$ & 7.62 & $-$3.96 & ($-$7.23)\\
         &         &          & 6158.149 & 10.741 & $-1.891$ & 7.62 & $-$3.96 & ($-$7.23)\\
         &         &          & 6158.172 & 10.741 & $-1.031$ & 7.62 & $-$3.96 & ($-$7.23)\\
         &         &          & 6158.187 & 10.741 & $-0.441$ & 7.61 & $-$3.96 & ($-$7.23)\\ 
\hline
\end{tabular}
\end{center}
Following columns 3--5 (laboratory wavelength, lower excitation potential, 
and $gf$ value), three kinds of damping parameters are presented in columns 6--8:  
Gammar is the radiation damping width (s$^{-1}$) [$\log\gamma_{\rm rad}$], 
Gammas is the Stark damping width (s$^{-1}$) per electron density (cm$^{-3}$) 
at $10^{4}$ K [$\log(\gamma_{\rm e}/N_{\rm e})$], and
Gammaw is the van der Waals damping width (s$^{-1}$) per hydrogen density 
(cm$^{-3}$) at $10^{4}$ K [$\log(\gamma_{\rm w}/N_{\rm H})$]. \\
All the data were taken from Kurucz and Bell (1995), except for 
the parenthesized damping parameters (unavailable in their compilation), 
for which the default values computed by the WIDTH9 program were assigned.
\end{table}

\end{document}